\documentclass[10pt, conference, letterpaper]{IEEEtran}
\newcommand*{\TechReport}{}%
\newcommand*{\IEEEHeadingFormat}{}%
\newcommand*{\AddSpace}{}%

\usepackage{comment}
\usepackage{array}
\usepackage{amsthm}
\usepackage[cmex10]{amsmath}
\usepackage{clrscode}
\usepackage[hidelinks]{hyperref}
\usepackage{graphicx}
\usepackage{enumerate}
\usepackage{color}
\usepackage{subcaption}
\usepackage{cite}
\usepackage{url}
\usepackage{amssymb}
\usepackage{multirow}

\newlength{\grafflecm}
\newcommand{\onec}{OneClock}

\IEEEoverridecommandlockouts
\ifdefined\TechReport
\else
\IEEEpubid{\makebox[\columnwidth]{978-1-5090-0223-8/16/\$31.00~\copyright~2016 IEEE \hfill} \hspace{\columnsep}\makebox[\columnwidth]{ }}
\fi

\begin{document}
\interfootnotelinepenalty=10000

\date{}

\title{
\ifdefined\TechReport
\onec\ to Rule Them All: \\ 
\vspace{3mm}
\hspace{7mm}
Using Time in Networked Applications 
\newline
\newline
				\large Technical Report\textsuperscript{\ensuremath\diamond}\thanks{\textsuperscript{\ensuremath\diamond}This technical report is an extended version of~\cite{OneClockNOMS}, which was accepted to IEEE/IFIP NOMS 2016.}, January 2016
\else
\onec\ to Rule Them All: \\ Using Time in Networked Applications 
\fi
}

\author{
{Tal Mizrahi, Yoram Moses\textsuperscript{\ensuremath*}\thanks{\textsuperscript{\ensuremath*}\scriptsize Yoram Moses is the Israel Pollak academic chair at Technion.}}\\
Technion --- Israel Institute of Technology
} 

\maketitle


\begin{abstract}
This paper introduces \onec, a generic approach for using time in networked applications. \onec\ provides two basic time-triggered primitives: the ability to schedule an operation at a remote host or device, and the ability to receive feedback about the time at which an event occurred or an operation was executed at a remote host or device. We introduce a novel prediction-based scheduling approach that uses timing information collected at runtime to accurately schedule future operations. 

Our work includes an extension to the Network Configuration protocol (NETCONF), which enables \onec\ in real-life systems. This extension has been published as an Internet Engineering Task Force (IETF) RFC, and a prototype of our NETCONF time extension is publicly available as open source.

Experimental evaluation shows that our prediction-based approach allows accurate scheduling in diverse and heterogeneous environments, with various hardware capabilities and workloads. \onec\ is a generic approach that can be applied to any managed device: sensors, actuators, Internet of Things (IoT) devices, routers, or toasters. 
\end{abstract}

\ifdefined\AddSpace\vspace{5mm}\fi
\section{Introduction}
\ifdefined\CutSpace \vspace{-2mm} \fi
\subsection{Background}
\textbf{Motivation.} Various distributed applications require the use of accurate time, including industrial automation systems~\cite{harris2008application}, automotive networks~\cite{IEEETSN}, and accurate measurement~\cite{moreira2009white}. Surprisingly, while these different applications typically use standard time synchronization methods (e.g., ~\cite{IEEE1588}), there is no standard method for \textbf{using time}, and thus each of these applications uses a proprietary management protocol that invokes time-triggered operations. In this paper we present a generic approach that allows the use of accurate time to manage various diverse devices, from routers to toasters.\footnote{Paraphrasing the 25-year old gimmick of the network-managed toaster~\cite{IntToaster}.}

\textbf{Why NETCONF?} 
\ifdefined\TechReport
A formal announcement by the Internet Engineering Steering Group (IESG), released in March 2014~\cite{WritableMIB}, declared that the IETF is encouraging the use of NETCONF~\cite{netconf}, rather than the Simple Network Management Protocol (SNMP)~\cite{snmp}. 
Indeed, the networking community is quickly shifting from SNMP-based Management Information Bases (MIB) to modules based on YANG~\cite{yang}, the modeling language used by NETCONF. During the writing of this paper, the IETF Active Internet Draft list~\cite{ActiveID} consisted of 256 drafts that define YANG data models~\cite{YangStat}, and only 21 drafts that define MIBs.
\else
A formal announcement by the Internet Engineering Steering Group (IESG), released in March 2014~\cite{WritableMIB}, declared that the IETF is encouraging the use of NETCONF~\cite{netconf}, rather than the Simple Network Management Protocol (SNMP)~\cite{snmp}. Indeed, the networking community is quickly shifting from SNMP-based Management Information Bases (MIB) to modules based on YANG~\cite{yang}, the modeling language used by NETCONF. 
\fi

\ifdefined\MgtText
\fi

NETCONF and YANG are gaining momentum in the context of various diverse applications, not only in the traditional realm of routers and switches, but also in other applications, such as Virtualized Network Functions~\cite{SFCYang} and Internet of Things (IoT) devices~\cite{sehgal2012management,manageIoT}. 
\ifdefined\TechReport
NETCONF is being adopted not only by the IETF, but also by other organizations, such as the Open Networking Foundation~\cite{OfConfig1.2}, and the Metro Ethernet Forum~\cite{MEF38}. 
\fi
\ifdefined\MgtText
In practice, NETCONF is gradually starting to realize the 25-year old gimmick of the network-managed toaster~\cite{IntToaster}.
\fi

We chose to use NETCONF as a baseline for \onec, due to its increasing adoption rate and diversity of applications and environments.

\ifdefined\AddSpace\vspace{3mm}\fi
\subsection{The \onec\ Protocol}
\label{OneProtocolSec}
\ifdefined\MgtText
In this paper we argue that time-triggered operations are essential in network management, especially in the context of NETCONF, due to its increasing adoption and diversity of applications.
\fi
In this paper we introduce a generic protocol for \textbf{using~time} in networked applications. The protocol is defined as an extension of NETCONF. A full specification of this extension, including an open-source YANG module that defines the extension, has been published as an RFC~\cite{NetconfTime}.

Our \onec\ extension defines two basic time-related primitives: (i) \textbf{scheduling}: a NETCONF client\footnote{We follow the NETCONF terminology; managed devices, referred to as \textbf{servers}, are managed by one or more \textbf{client}.} can schedule a Remote Procedure Call (RPC) to be performed by a NETCONF server at a prescribed future time, and (ii) \textbf{reporting}: a NETCONF client can receive feedback about the time of execution of an RPC, or a notification about the time of occurrence of a monitored event.


\ifdefined\TechReport
\onec\ can be used in various important use cases, such as invoking scheduled operations in diverse applications, taking coordinated snapshots of a system, or performing network-wide atomic commits. 
\fi

\ifdefined\AddSpace\vspace{3mm}\fi
\subsection{\onec: Accurate Scheduling}
One of the greatest challenges in our approach is to \emph{accurately} schedule network operations. Even if a managed device (server) keeps an accurate clock, it is difficult to guarantee that scheduled operations are performed very close to their scheduled times. The actual execution time may depend on the processing power of the server, on its operating system, and its load due to other tasks that run in parallel.

We propose a prediction-based approach that allows a client to accurately schedule network operations without prior knowledge about the servers' performance. The approach is based on measuring the Elapsed Time of Execution (ETE) of each RPC, and using previous ETE measurements to predict the next ETE.

\begin{figure}[htbp]
	\ifdefined\CutSpace \vspace{-1mm} \fi
  \centering
  \fbox{\includegraphics[height=2.75\grafflecm]{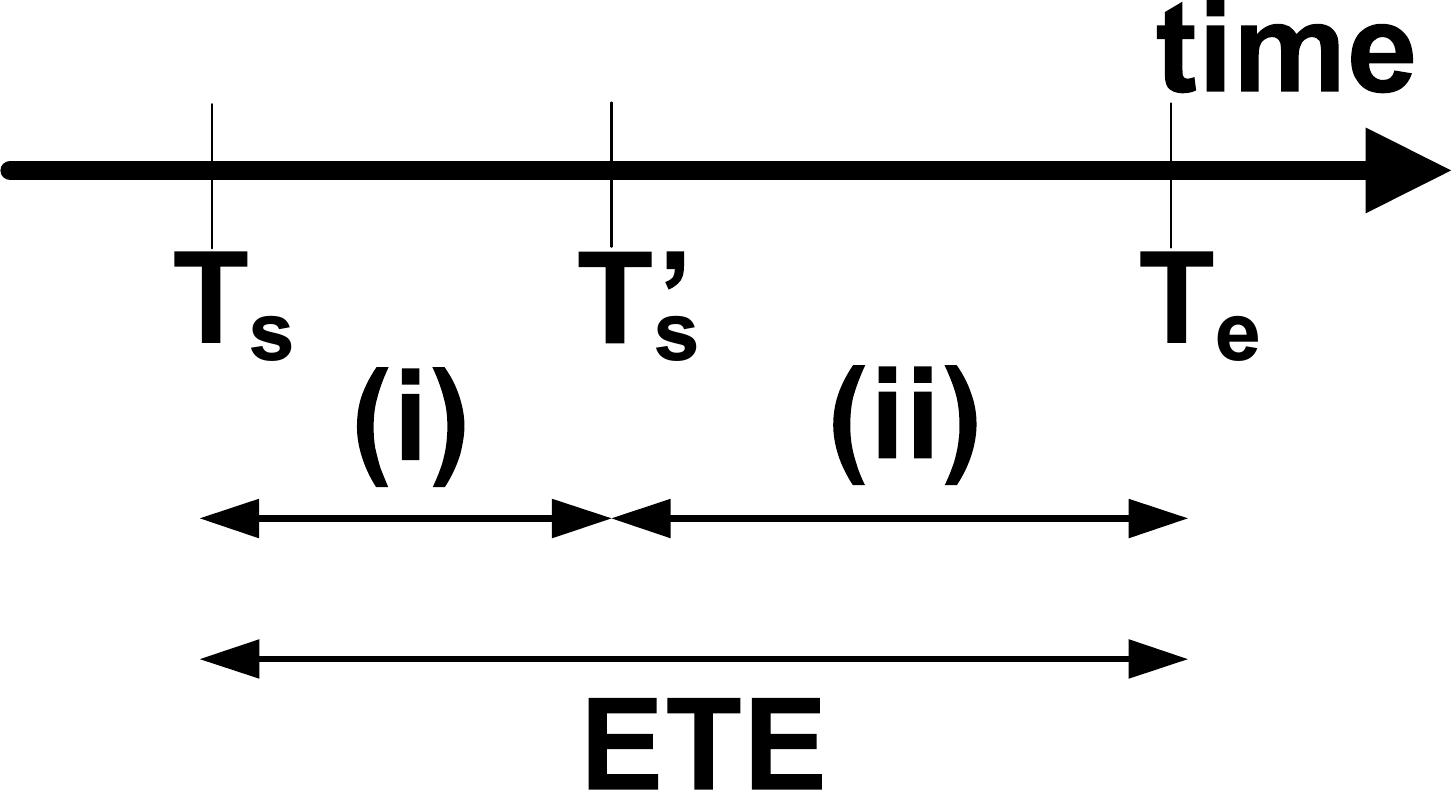}}
  \caption{Elapsed Time of Execution (ETE): ETE~=~${T_e-T_s}$.}
  \label{fig:ETE}
\end{figure}

\begin{figure}[htbp]
  \centering
  \fbox{\includegraphics[width=.47\textwidth]{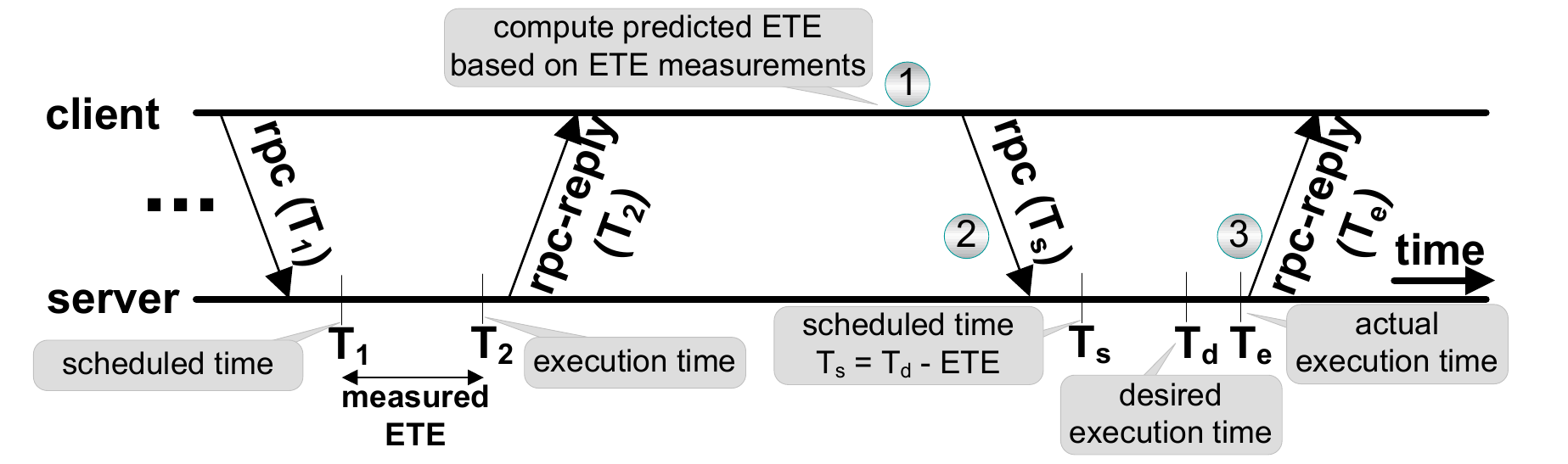}}
  \caption{Prediction-based scheduling: by predicting the ETE, a client can control when the RPC will be \emph{completed}.}
  \label{fig:Prediction}
\end{figure}

The ETE is defined to be $T_e-T_s$ (see Fig.~\ref{fig:Prediction}), where $T_s$ is the scheduled \emph{start time} of the RPC, and $T_e$ is the actual \emph{completion time} of the RPC. The \emph{actual} start time of the RPC is denoted by $T'_s$. Hence, as depicted in Fig.~\ref{fig:ETE}, the ETE is affected by two non-deterministic factors: (i) the server's ability to accurately start the operation, and (ii) the running time of the RPC. 

For each scheduled operation (see the numbered steps in Fig.~\ref{fig:Prediction}):\footnote{We follow the notation of~\cite{netconf}, where Remote Procedure Calls are denoted by uppercase RPC, and the messages that carry RPCs are denoted by lowercase \emph{rpc}.}

\begin{sloppypar}
\begin{enumerate}
	\item The client predicts the ETE of the next RPC based on previous measurements of the \emph{scheduled time} and \emph{execution time}. 
	\item For a given desired execution time, $T_d$, the client schedules the operation to be performed at~${T_d-ETE}$.
	\item The server reports the \emph{actual time of execution}, $T_e$, back to the client, allowing the client to use this feedback for scheduling future operations.
\end{enumerate}
\end{sloppypar}

Notably, our scheduling approach allows a NETCONF client to accurately schedule network operations in a heterogeneous environment, where the performance of the managed servers is not necessarily known in advance.


\ifdefined\AddSpace\vspace{3mm}\fi
\subsection{Related Work}

The use of the time-of-day in network management is a common practice. Time-of-day routing~\cite{ash1985use} routes traffic to different destinations based on the time-of-day. Scheduled operations~\cite{rfc3231,conditional} allow various policies and configurations to be applied at specific time ranges. 
The work of~\cite{TimedConsistent} analyzed the use of timed path updates in Software Defined Networks. This paper introduces a more general framework that allows time-triggered operations in any network managed device, and enables accurate scheduling of network operations in a heterogeneous environment. The work of~\cite{InfocomTimeFlip} suggested a method for accurate scheduling in switches and routers using Ternary Content Addressable Memories (TCAM). Our scheduling scheme is more generic, as it makes no assumption about the hardware of the managed devices.

The literature is rich with works that analyze and predict program running times, e.g.,~\cite{IversonOP99,IpekSSM05,GiustoMH01,IversonOF96,BontempiK02}. In this paper we use \emph{time series analysis} to predict the execution time of a \emph{remote operation}, allowing to perform \emph{accurate scheduling} of the requested operation. 

\ifdefined\AddSpace\vspace{3mm}\fi
\subsection{Contributions}
The main contributions of this paper are:

\begin{itemize}
	\item We introduce \onec, a generic approach for using time in networked applications. \onec\ defines two basic primitives, \emph{schedule}, and \emph{report}. Several use cases that demonstrate the merits of \onec\ are presented.
	\item We present a scheduling approach that allows accurate scheduling by predicting the server's execution time. We analyze three prediction algorithms: two average-based algorithms, and a Kalman-Filter-based algorithm.
	\item We define a \onec\ extension to NETCONF, which has been published as an IETF RFC. 
	\item We have implemented a prototype of the NETCONF time extension. Our prototype is available as open source. Our experimental evaluation demonstrates how accurately events can be scheduled over a network.
\end{itemize}

\ifdefined\TechReport
\else
Due to space limits, some of the details of our analysis are presented in a technical report~\cite{OneClockTR}.
\fi

\ifdefined\AddSpace\vspace{5mm}\fi
\section{Using \onec\ in Practice}
In this section we describe three use cases that illustrate how the two time-triggered primitives, \emph{scheduling} and \emph{reporting}, can be used in distributed systems.

\ifdefined\AddSpace\vspace{3mm}\fi
\subsection{Coordinated Operation}
It is often desirable to coordinate a set of events or operations that should take place at different nodes in the system at the same time,\footnote{In practical systems it is typically not possible to coordinate events to be performed \emph{exactly} at the same time at different nodes. Throughout the paper, the term `same time' should be read as `same time within the accuracy limitations of the servers'.} or should occur according to a specific order. The \emph{schedule} primitive can be used to coordinate events occurring in actuators in a factory product line~\cite{harris2008application}, to coordinate a routing change in a network~\cite{I2RSreq}, or to orchestrate events in scientific experiments~\cite{cernWhiteRabbit}.

Using \onec, a client can \emph{schedule} a simultaneous event at multiple servers, or define a sequence of scheduled times that determine the order and relative timing of events.

\begin{figure}[htbp]
	\centering
  \begin{subfigure}[t]{.24\textwidth}
  \centering
  \fbox{\includegraphics[height=3.3\grafflecm]{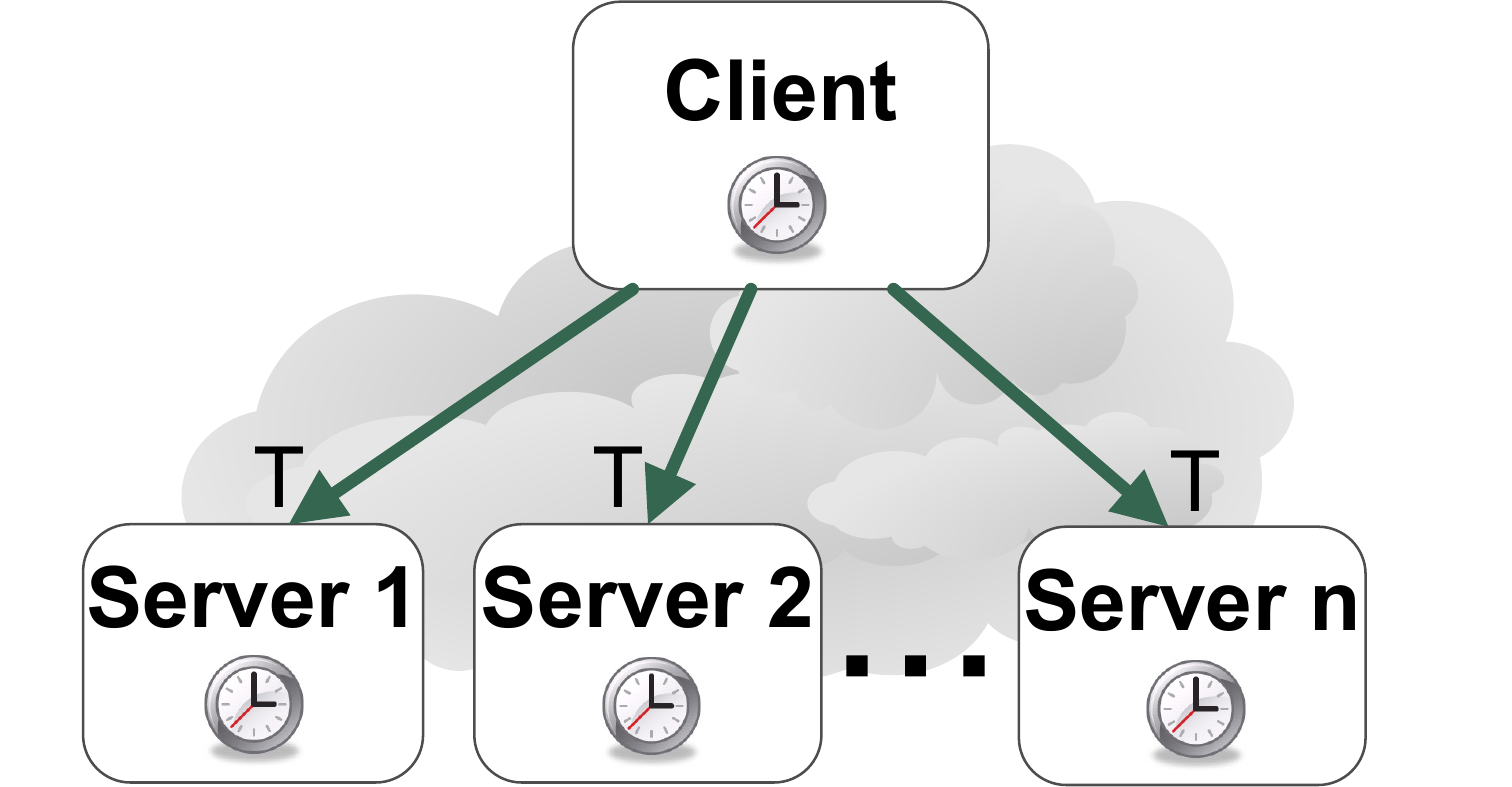}}
  \caption{Coordinated operations: all servers perform the operation at the same time, $T$.}
  \label{fig:Coordinated}
  \end{subfigure}%
  \begin{subfigure}[t]{.25\textwidth}
  \centering
  \fbox{\includegraphics[height=3.3\grafflecm]{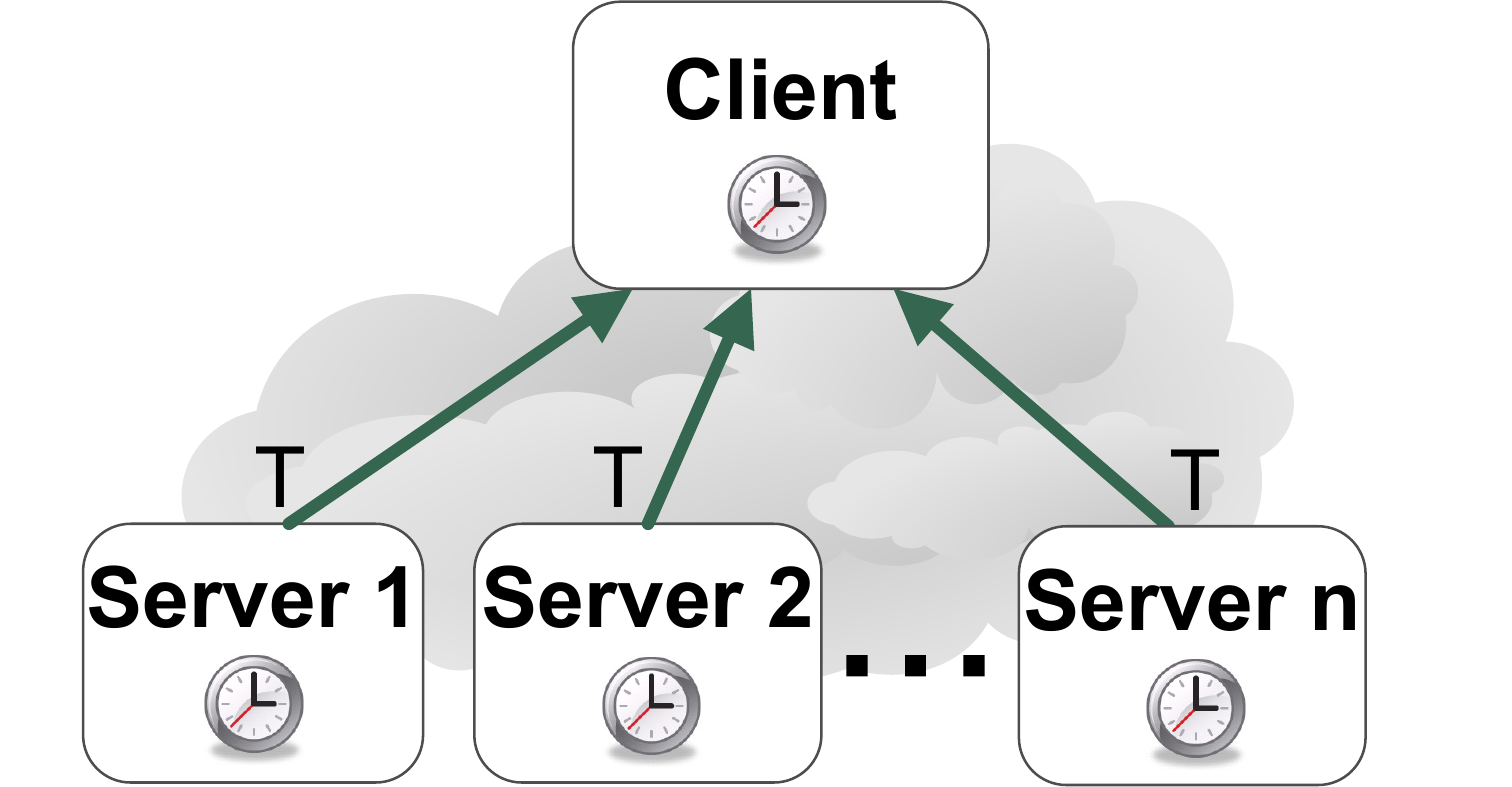}}
	\captionsetup{justification=centering}
  \caption{Coordinated snapshot: all servers send their state to the client at the same time.}
  \label{fig:Snapshot}
  \end{subfigure}%
  \caption{Coordinated operations and coordinated snapshots.}
  \label{fig:UseCases}
\end{figure}

\begin{figure*}[htbp]

	\centering
  \begin{subfigure}[t]{.25\textwidth}
  \centering
  \fbox{\includegraphics[height=3.0\grafflecm]{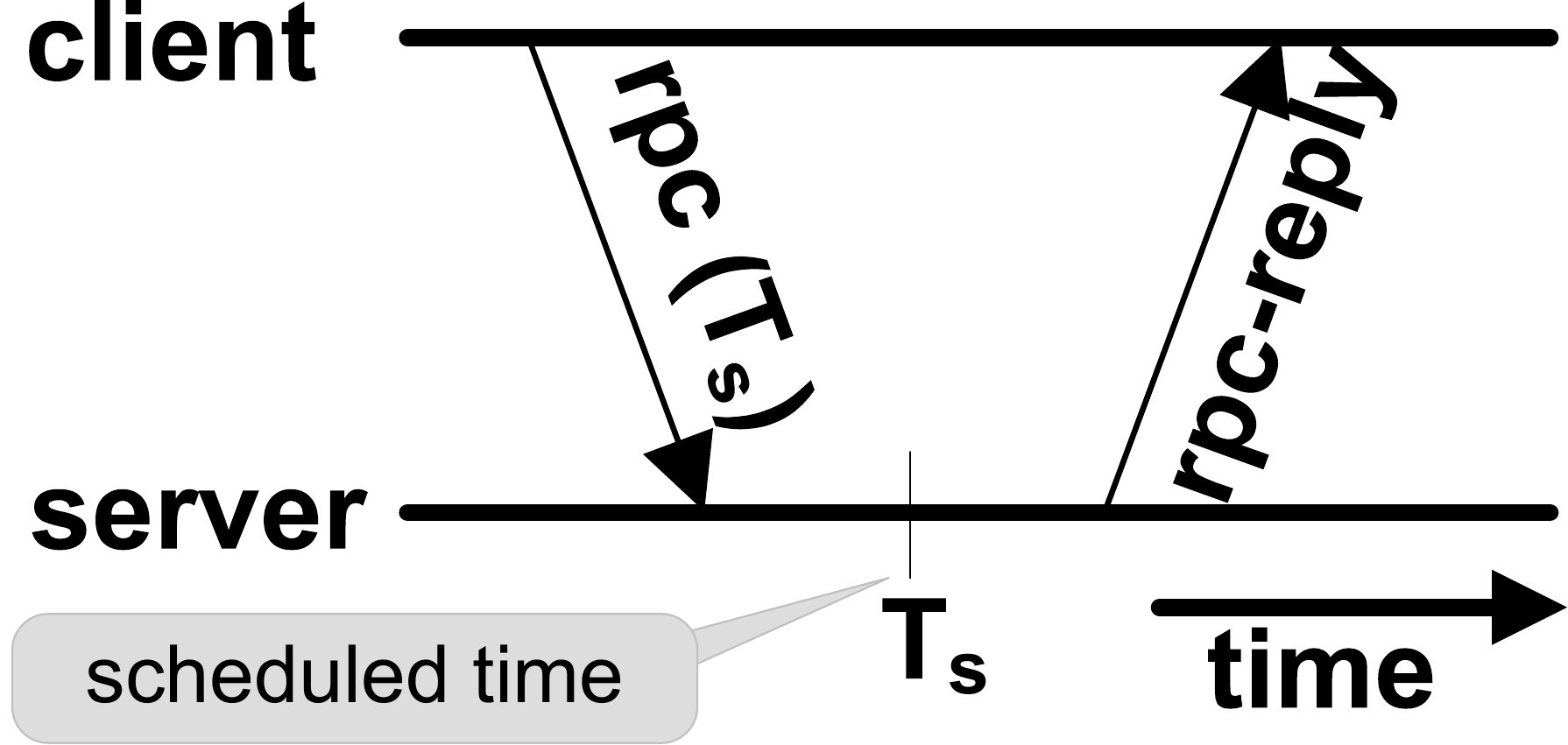}}
	\captionsetup{justification=centering}
  \caption{Scheduled RPC.}
  \label{fig:ScheduledRPC}
  \end{subfigure}%
  \begin{subfigure}[t]{.25\textwidth}
  \centering
  \fbox{\includegraphics[height=3.0\grafflecm]{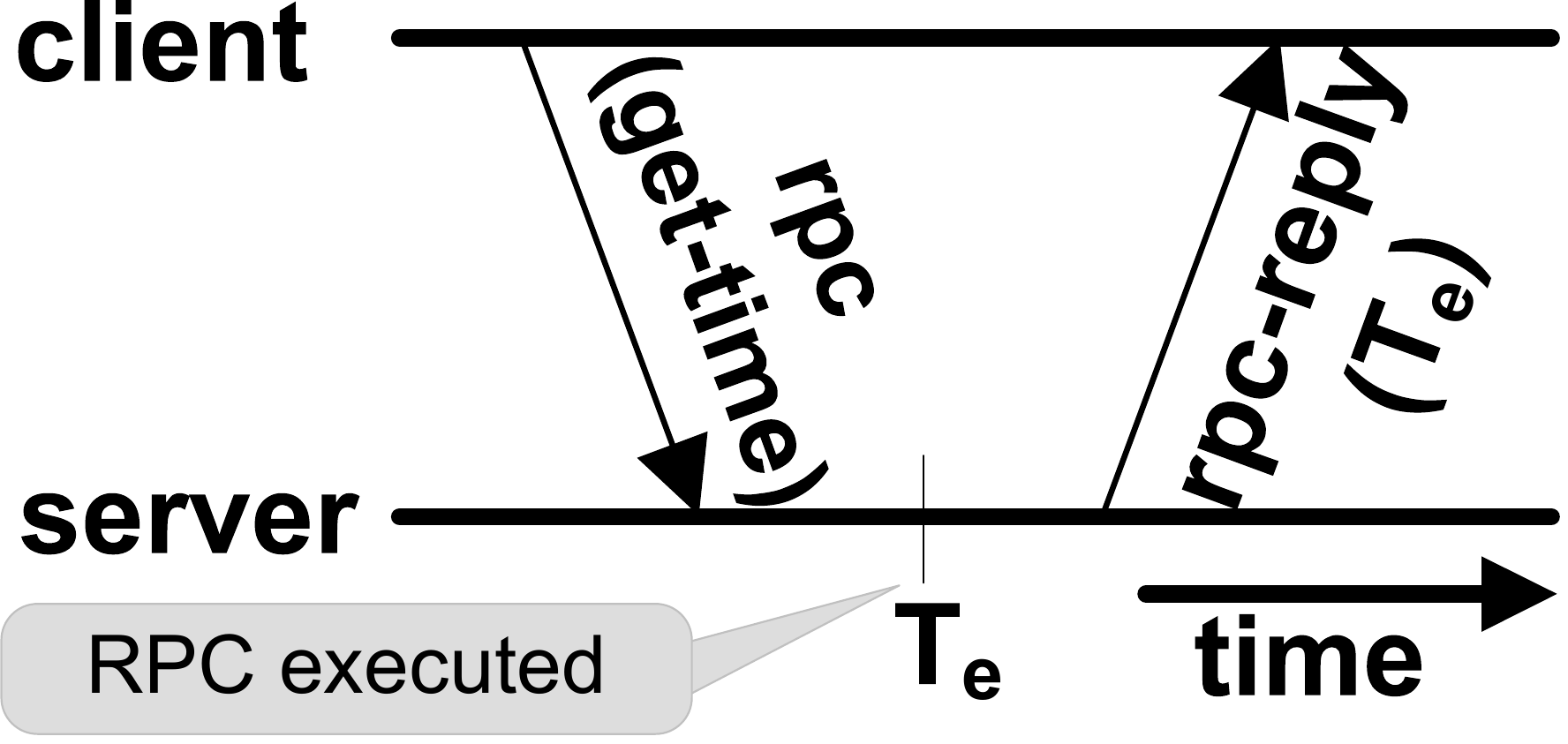}}
	\captionsetup{justification=centering}
  \caption{Reporting the execution time.}
  \label{fig:ReportingRPC}
  \end{subfigure}%
  \begin{subfigure}[t]{.25\textwidth}
  \centering
  \fbox{\includegraphics[height=3.0\grafflecm]{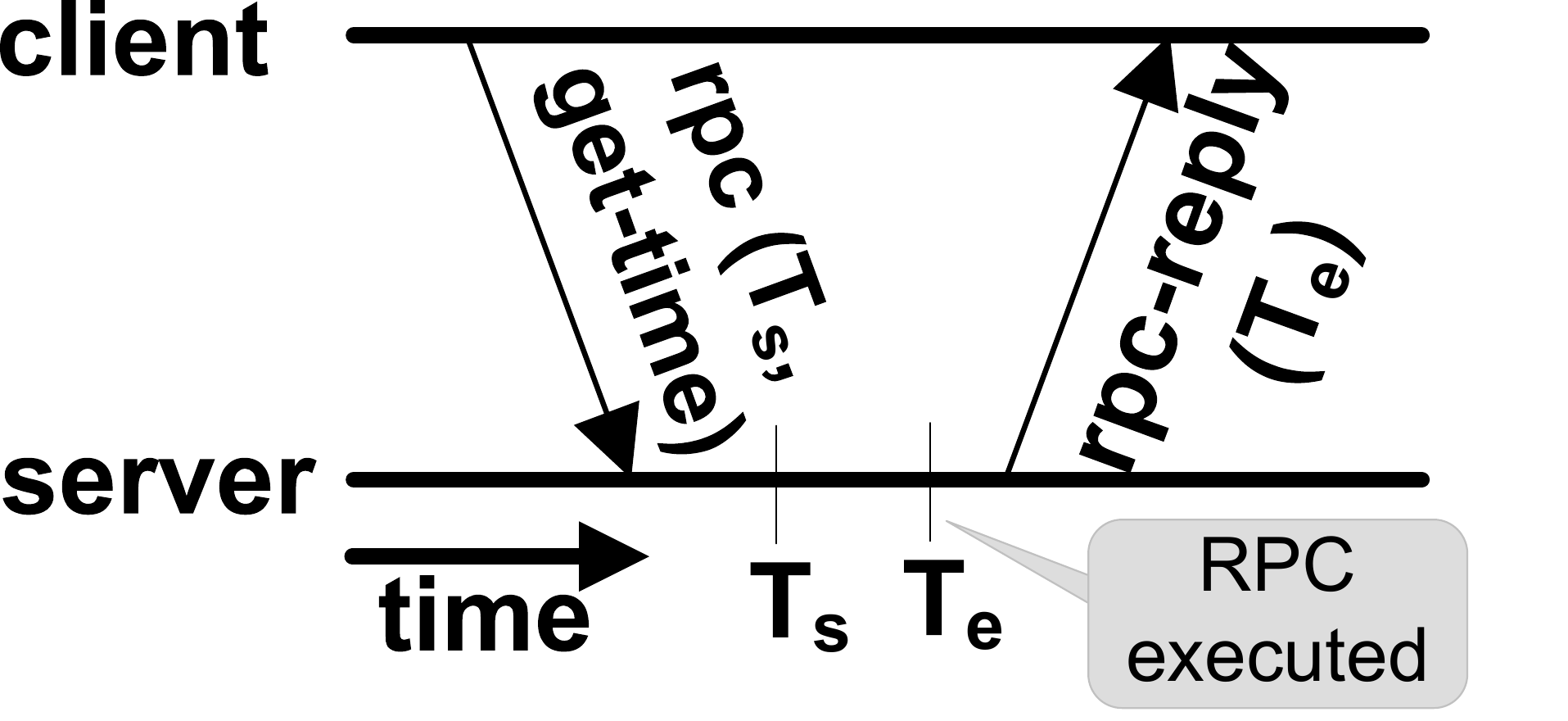}}
	\captionsetup{justification=centering}
  \caption{Reporting the execution time of a scheduled RPC.}
  \label{fig:SchReportingRPC}
  \end{subfigure}%
  \begin{subfigure}[t]{.25\textwidth}
  \centering
  \fbox{\includegraphics[height=3.0\grafflecm]{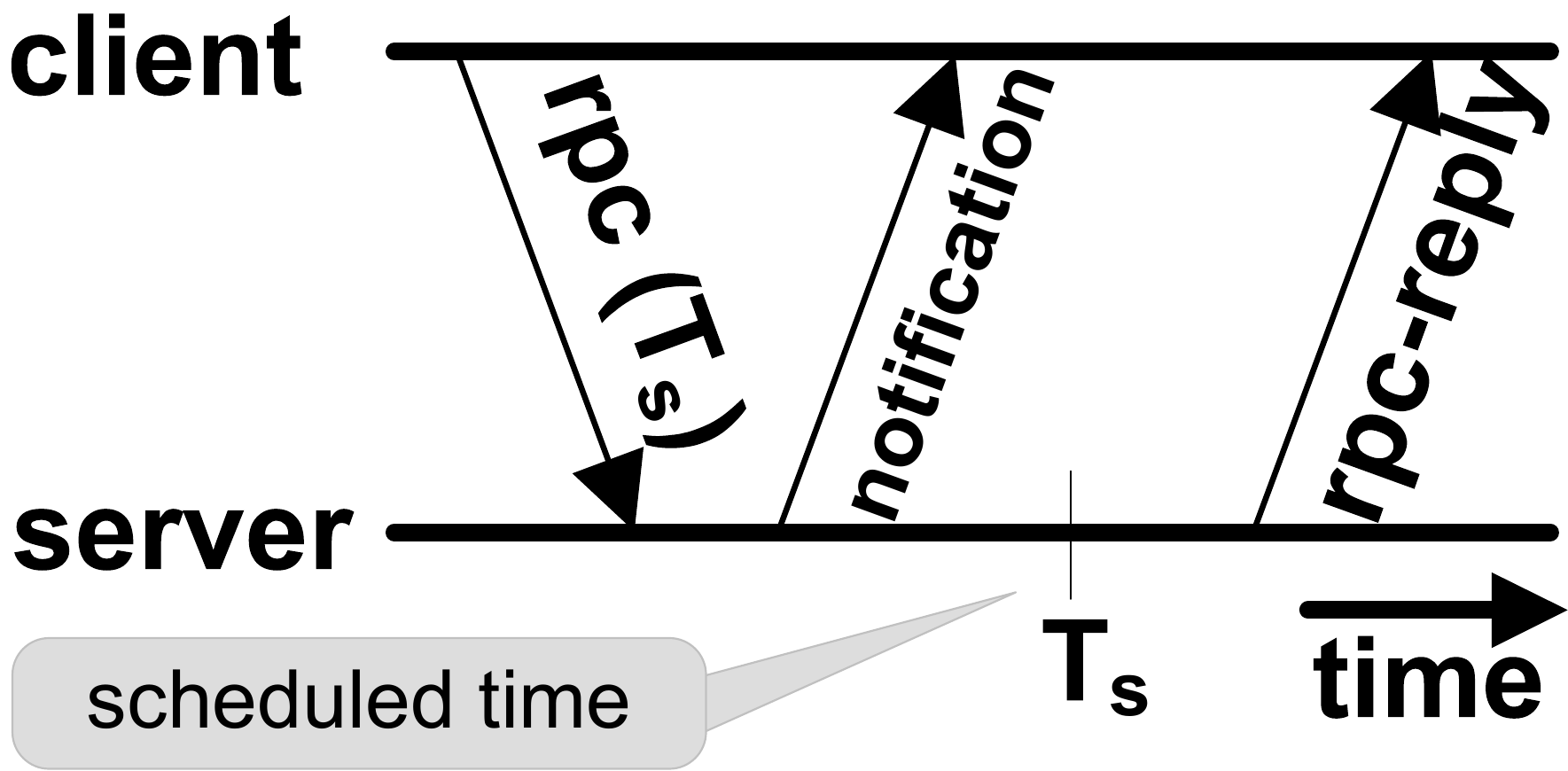}}
	\captionsetup{justification=centering}
  \caption{Scheduled RPC with notification.}
  \label{fig:ScheduledRPCNoti}
  \end{subfigure}%
  \caption{The time capability in NETCONF.}
  \label{fig:NETCONFext}
	\ifdefined\CutSpace \vspace{-2mm} \fi
\end{figure*}

\ifdefined\AddSpace\vspace{3mm}\fi
\subsection{Coordinated Snapshot}
In many applications it is desirable to monitor events or statistics with respect to a common time reference. 

A client can perform a \emph{coordinated snapshot}, i.e., capture the state of a monitored attribute at all the servers at the same time. While a simultaneous snapshot does not produce a consistent distributed snapshot~\cite{ChandyL85}, it provides a coordinated snapshot of the servers' state. For example, when collecting statistics from all the servers, it is most useful to capture the information at the same time in all servers. In power grid networks~\cite{dickerson2010time}, synchrophasor measurements are used for monitoring the operation of a power grid. These measurements must be synchronized, so as to allow correct system-wide processing.

\begin{sloppypar}
\onec\ enables coordinated snapshots; using the \emph{schedule} primitive, a client can schedule a NETCONF \emph{get-config} operation~\cite{netconf} to be taken at time $T$, causing all the servers to send their response at the same time (Fig.~\ref{fig:Snapshot}). 
\end{sloppypar}

\ifdefined\AddSpace\vspace{3mm}\fi
\subsection{Network-wide Atomic Commit}
\label{AtomicCommitSec}
\ifdefined\TechReport

\begin{figure*}[htbp]
  \centering
  \fbox{\includegraphics[width=27\grafflecm]{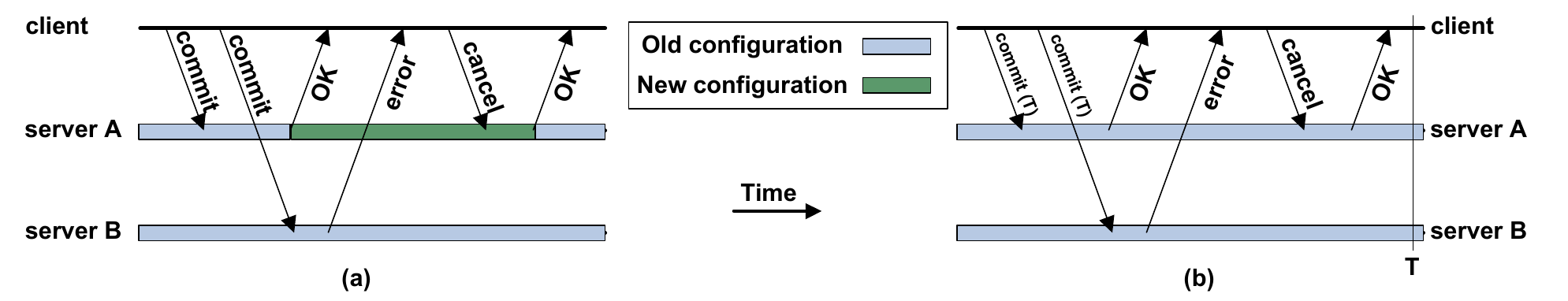}}
  \caption{Atomic commit: (a) NETCONF confirmed commit, without using time. (b) Time-triggered commit.}
  \label{fig:ConfirmedCommit}
\end{figure*}

The NETCONF \emph{commit}~\cite{netconf} is an RPC that commits the \emph{candidate} configuration, i.e., copies the \emph{candidate} configuration to the \emph{running} configuration. This operation allows the client to prepare a set of configuration updates in the \emph{candidate} datastore, and then apply them at once with the \emph{commit} operation.

It is often desirable to perform a \emph{network-wide atomic commit}, where either all the servers successfully perform the commit operation, or if some of the servers are not able to perform the commit, then none of the servers perform it.

Atomic commits can be performed using NETCONF without our time extension, but potentially at the cost of a temporary state of inconsistency, where different servers use different configuration versions (Fig.~\ref{fig:ConfirmedCommit}). This can be done using the NETCONF \emph{confirmed commit} procedure. This procedure requires two steps: (i) the client sends a first commit message to all the servers, causing them to switch to the candidate configuration, and (ii) the client sends a \emph{confirming commit} message to all the servers, finalizing the commit procedure. If the two phases are not completed successfully, or if the client cancels the commit, the servers roll back to the previous configuration.

The \onec\ \emph{schedule} primitive enables a clean and straightforward approach to network-wide commits, as illustrated in Fig.~\ref{fig:ConfirmedCommit}. The client sends a scheduled commit message to the servers, to be performed at a future time $T$. If some of the servers fail to schedule the commit operation, the client can cancel the commit before time $T$, leaving all the servers at the current configuration.  
\else
The \emph{schedule} primitive allows a client to perform a network-wide commit. The client can schedule the commit operation to a time in the future, and if one of the servers indicates that it will not be able to perform the operation, the client can send a cancellation message to the other servers, as described in Sec.~\ref{NotifCancelSec}, thereby guaranteeing atomic operation.
\fi


\ifdefined\AddSpace\vspace{5mm}\fi
\section{NETCONF Time Extension}
We introduce an extension to the NETCONF protocol that allows time-triggered operations. 
\ifdefined\TechReport
The extension is defined as a new \emph{capability}~\cite{netconf}. 
\fi
\ifdefined\BlindReview
Details are presented in an anonymous specification~\cite{TimeCap}. 
\else
Details are presented in~\cite{NetconfTime}. 
\fi

\subsection{Overview}
\label{ExtOverviewSec}
The time capability provides two main functions:
\begin{itemize}
	\item \textbf{Scheduling.} When a client sends an \verb|rpc| message to a server, the message may include the \verb|scheduled-time| parameter, denoted by $T_s$ in Fig.~\ref{fig:ScheduledRPC}. The server then starts to execute the RPC as close as possible to the scheduled time $T_s$, and once completed the server can respond with an \verb|rpc-reply| message.
	\item \textbf{Reporting.} When a client sends an \verb|rpc| message to a server, the message may include a \verb|get-time| element (see Fig.~\ref{fig:ReportingRPC}), requesting the server to return the execution time of the RPC. In this case, after the server performs the RPC it responds with an rpc-reply that includes the \verb|execution-time| parameter, specifying the time $T_e$ at which the RPC was completed.
\end{itemize}

The two scenarios discussed above imply that a third scenario can also be supported (Fig.~\ref{fig:SchReportingRPC}), where the client sends an \verb|rpc| message that includes a scheduled time, $T_s$, as well as the \verb|get-time| element. This allows the client to receive feedback about the actual execution time, $T_e$. Ideally, $T_s=T_e$. However, the server may execute the RPC at a slightly different time than $T_s$, for example if the server is tied up with other tasks at time $T_s$.

The \emph{report} abstraction, presented in Sec.~\ref{OneProtocolSec}, allows the client to receive information about the execution time of an RPC, or to receive notifications about the time of occurrence of events. The former can be implemented using the \verb|get-time| procedure we defined, while the latter is already supported in NETCONF by using notifications that include the \verb|eventTime| parameter~\cite{netconfnotifications}. 

\ifdefined\AddSpace\vspace{3mm}\fi
\subsection{Applying the Time Primitives to Various Applications}
The time capability specification 
\ifdefined\BlindReview
we defined~\cite{TimeCap} 
\else
we defined~\cite{NetconfTime} 
\fi
includes a YANG module that adds the two new primitives, \emph{schedule} and \emph{report}, as two parameters in all the RPC types defined in~\cite{netconf}. 
\ifdefined\TechReport
For example, this YANG module enables scheduled \verb|commit|, and scheduled \verb|set-config| RPCs.
\fi

Notably, the time primitives are not limited to the RPCs defined in~\cite{netconf}. If a new YANG module defines a new RPC, the module can include the time parameters, allowing the new RPC to use the time primitives. Our open source code includes two such examples:

\begin{itemize}
	\item We enhanced the well-known \emph{toaster} YANG module~\cite{ToasterYANG}, by allowing the \verb|make-toast| operation to be a scheduled RPC.
	\item We created a new YANG module called \emph{test}, which triggers the server to perform a configurable command line. Using the \emph{schedule} parameter, the \emph{test} RPC can be used as a remote variant of the well-known Cron~\cite{cron} command in Linux.
\end{itemize}

The \emph{report} primitive can be used not only by applying the \verb|get-time| parameter, but also by other means that are inherently possible when using NETCONF. The time of occurrence of important events can be sent to the client using a NETCONF notification~\cite{netconfnotifications}, or can be included in the NETCONF data model. For example, the YANG data model that defines a log entry may include the time-of-day in each log entry.

\ifdefined\AddSpace\vspace{3mm}\fi
\subsection{Notifications and Cancellation Messages}
\label{NotifCancelSec}
\ifdefined\IEEEHeadingFormat
\vspace{2mm}
\textbf{1. Notifications}
\vspace{2mm}

\else
\subsubsection{Notifications}
\fi
As illustrated in Fig.~\ref{fig:ScheduledRPC}, after a scheduled RPC is executed the server sends an \verb|rpc-reply|. The \verb|rpc-reply| may arrive a long period of time after the \verb|rpc| message was sent by the client, leaving the client without a clear indication of whether the \verb|rpc| was received.
Therefore, we define an optional \verb|netconf-scheduled-message| notification (Fig.~\ref{fig:ScheduledRPCNoti}), which provides an immediate acknowledgment of the scheduled RPC. 
As illustrated in Fig.~\ref{fig:ScheduledRPCNoti}, when the server receives a scheduled RPC it sends a notification that includes the \verb|message-id| of the scheduled RPC.

\begin{figure}[htbp]
  \centering
  \fbox{\includegraphics[width=11\grafflecm]{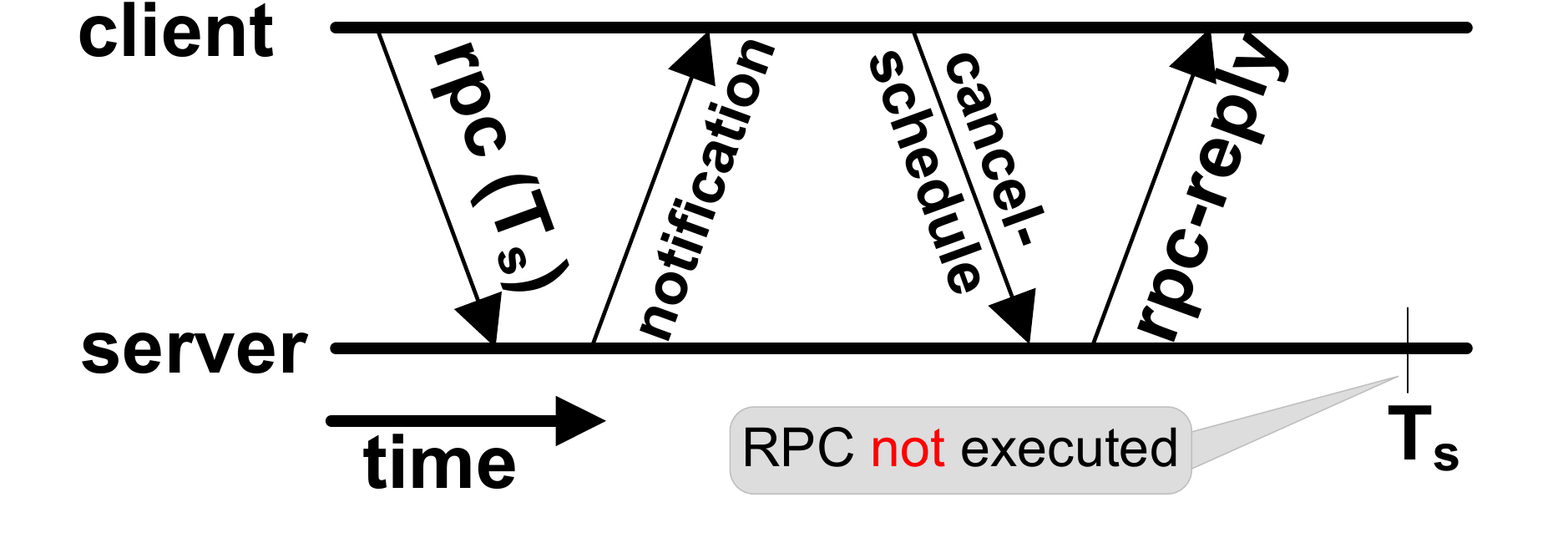}}
  \caption{Cancellation message.}
  \label{fig:ScheduledRPCCancel}
	\ifdefined\CutSpace \vspace{-3mm} \fi
\end{figure}

\ifdefined\IEEEHeadingFormat
\vspace{2mm}
\textbf{2. Cancellation Messages}
\vspace{2mm}

\else
\subsubsection{Cancellation Messages}
\fi
A client can cancel a scheduled RPC by sending a \verb|cancel-schedule| RPC (Fig.~\ref{fig:ScheduledRPCCancel}).
The \verb|cancel-schedule| RPC, defined in this document, can be used to enforce the coordinated network-wide commit described in Sec.~\ref{AtomicCommitSec}. 

\ifdefined\AddSpace\vspace{3mm}\fi
\subsection{Clock Synchronization}
The time capability we defined requires clients and servers to maintain clocks. It is assumed that clocks are synchronized 
\ifdefined\TechReport
by a clock synchronization method, e.g., ~\cite{mills2010rfc,IEEE1588}. 
\else
e.g., by~\cite{mills2010rfc,IEEE1588}.
\fi

\ifdefined\AddSpace\vspace{3mm}\fi
\subsection{Acceptable Scheduling Range}
\begin{figure}[!b]
	\ifdefined\CutSpace \vspace{-3mm} \fi
  \centering
  \fbox{\includegraphics[width=11\grafflecm]{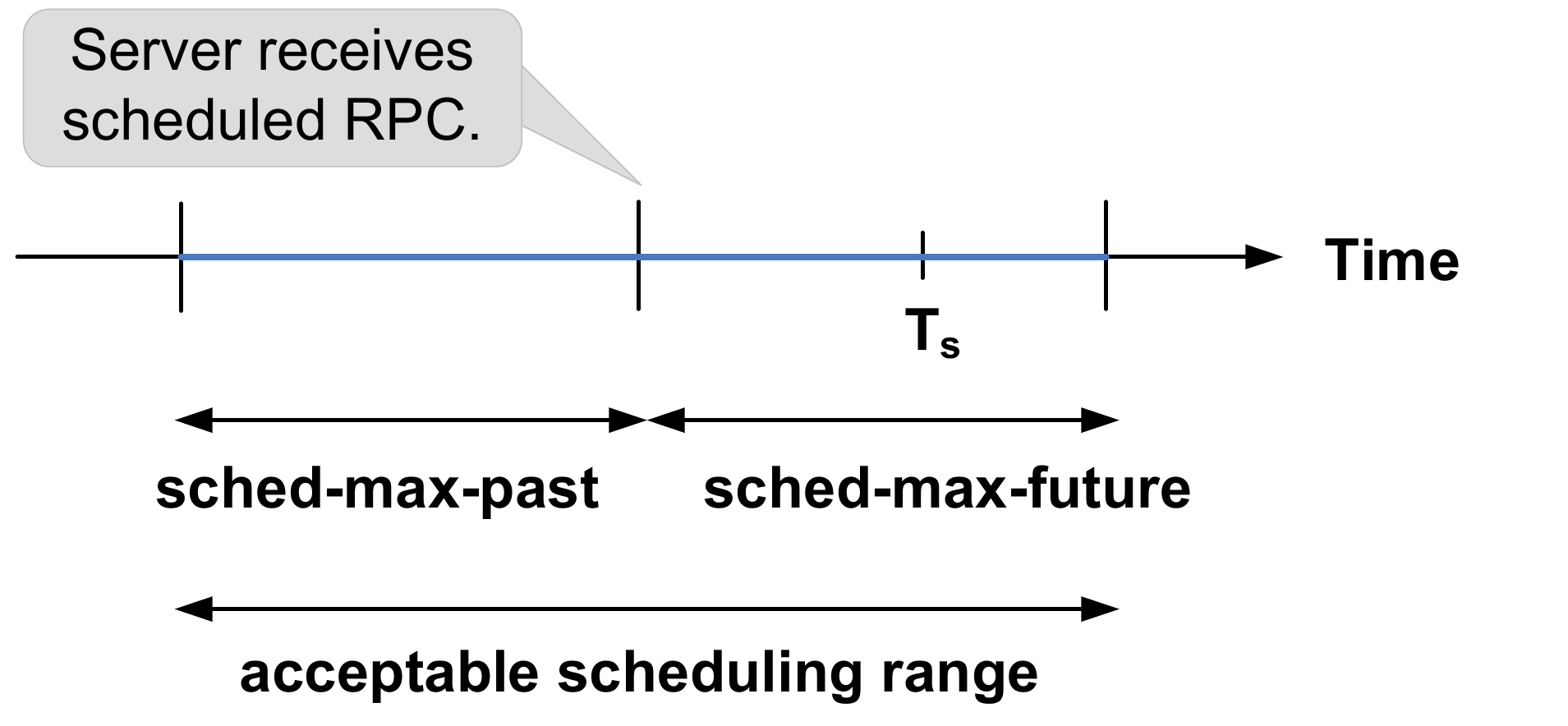}}
  \caption{Acceptable scheduling range: defined by two configurable parameters: sched-max-future and sched-max-past.}
  \label{fig:SchedRange}
\end{figure}

A server that receives a message that is scheduled to be performed at time $T_s$ verifies that the value $T_s$ is not too far in the past or in the future. As illustrated in Fig.~\ref{fig:SchedRange}, the server verifies that $T_s$ is within the acceptable scheduling range.

If $T_s$ occurs in the past and within the acceptable scheduling range, the server performs the RPC as soon as possible

The scheduling bound defined by \verb|sched-max-future| guarantees that every scheduled RPC is restricted to a near future scheduling time, on the order of seconds, and not on the order of hours or days. This restriction significantly reduces the impact of potential coherency problems that may result from server failures, or from multiple clients trying to schedule conflicting operations. 
\ifdefined\TechReport
\else
This issue is further discussed in~\cite{OneClockTR}.
\fi

\ifdefined\AddSpace\vspace{5mm}\fi
\section{Prediction-based Scheduling}
\label{PredictionSec}
Our scheduling approach is based on using previous measurements of the Elapsed Time of Execution (ETE). 

\begin{figure}[htbp]
	\centering
  \fbox{\includegraphics[height=2.75\grafflecm]{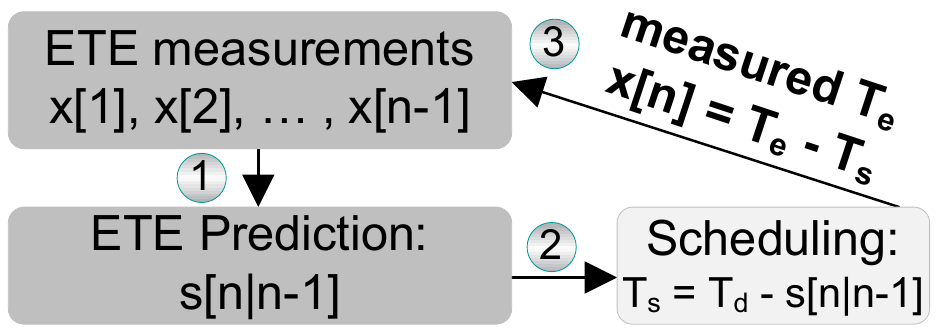}}
  \caption{Prediction-based scheduling approach.}
  \label{fig:PredictApproach}
	\ifdefined\CutSpace\vspace{-3mm}\fi
\end{figure}

Based on the ETE measurements, the client uses the prediction approach illustrated in Fig.~\ref{fig:PredictApproach}. The prediction approach consists of three steps:
\begin{enumerate}
	\item When a scheduled operation is required to take place at time $T_d$, the client uses previous ETE measurements, $x[1], \ldots, x[n-1]$, to predict the next ETE, denoted by $s[n|n-1]$.
	\item The next scheduled time is $T_s=T_d-s[n|n-1]$.
	\item The client updates its measurement set based on the feedback received from the server about the execution time $T_e$.
\end{enumerate}

In the rest of this section we describe the two main components of our scheduling approach, the ETE measurements, and the ETE prediction.

\ifdefined\AddSpace\vspace{3mm}\fi
\subsection{ETE Measurements}
\label{ETEMeasureSec}
We analyze two measurement methods:

\textbf{Periodic probing.} This approach uses ETE measurements that are taken periodically at a constant frequency. In systems that require periodic operations, these ETE measurements are inherently available. In other systems, the client can proactively send periodic scheduled RPCs to every server in order to probe the ETE. The main drawback of periodic probing is that it can potentially consume unnecessary resources, both at the client and at the server.

\textbf{Burst probing.} The second approach uses an on-demand burst of probe RPCs; when a scheduled RPC is required, the client initiates a burst of $N$ scheduled RPCs, performed at a fixed frequency. This approach does not require resource consumption in the absence of actual scheduled RPC requests, but the prediction is potentially less accurate, since it is based on a smaller number of measurements.

Note that both approaches require the probe RPCs to be similar in terms of performance and running time to the future RPC for which the prediction is required.



\ifdefined\TechReport
\else
Further analysis of measurement periods is provided in~\cite{OneClockTR}.
\fi

\ifdefined\MeasurementPeriodText
The measurement period selection policy should be slightly different for burst probing and for periodic probing. In burst probing, we want the burst to take as little time as possible, and therefore we would like to the period to be as short as possible. In periodic probing we would like to avoid loading the server's resources, and thus we would like the period to be as high as possible, without reducing the scheduling accuracy.

For burst probing, we propose a simple period selection approach, which should be performed after a client-server session is initialized:

\begin{enumerate}
	\item The client sends $M$ probe bursts to the server, where each burst consists of $N$ RPCs. 
	\item $P_i = 2 P_{i-1}$ for each probe $i$, where $2 \leq i \leq M$, and the period of burst $i$ is denoted by $P_i$.
	\item The client computes the average ETE in each burst. Let $j$ be the burst with the minimal ETE.
	\item $P^B \gets 2 \cdot P_j$. 
\end{enumerate}

The period $P^B$, is chosen to be slightly higher than the optimal ETE, leaving a safety margin that allows the server to be more loaded than during the measurement without affecting the ETE.
	
For periodic probing we propose a similar selection algorithm, except that step~4 is replaced by:

$P^P \gets \max \limits_{1 \leq k \leq M} P_k < (1+\alpha) \cdot P_j$ for a previously chosen constant $0 < \alpha < 1$.

This approach guarantees that we select a long period that does not yield a significantly higher ETE than the optimal ETE.
\fi

\ifdefined\AddSpace\vspace{3mm}\fi
\subsection{ETE Prediction Algorithms}
We analyzed three prediction algorithms, \textbf{Average}, \textbf{FT-Average}, and \textbf{Kalman}. We now describe these algorithms.

\ifdefined\IEEEHeadingFormat
\vspace{2mm}
\textbf{1. Baseline}
\vspace{2mm}

\else
\subsubsection{Baseline}
\fi
The baseline for comparison in our evaluation is the simplest approach which assumes $s[n]=0$, and therefore assigns $T_s=T_d$. In this approach the prediction error is equal to the ETE.

\ifdefined\IEEEHeadingFormat
\vspace{2mm}
\textbf{2. Average Algorithm}
\vspace{2mm}

\else
\subsubsection{Average Algorithm}
\fi
The \textbf{Average} algorithm performs an average of the last $N$ measurements:
\ifdefined\CutSpace \vspace{-3mm} \fi

\begin{equation}
s[n|n-1]=\frac{1}{N} \sum \limits_{j=1}^{N} x[n-j]
\end{equation}

\ifdefined\IEEEHeadingFormat
\textbf{3. Fault-tolerant Average (FT-Average) Algorithm}
\vspace{2mm}

\else
\subsubsection{Fault-tolerant Average (FT-Average) Algorithm}
\fi
The Fault-tolerant Average~\cite{LundeliusL84} performs an average of the last $N$ measurement samples, after ignoring the highest and the lowest measurement values. Hence, this approach masks the most noisy or erroneous measurement of the $N$ samples.

\begin{equation}
\begin{split}
s[n|n-1]=
    \begin{cases}
      \frac{1}{N} \sum \limits_{j=1}^{N} x[n-j], & \text{if}\ N<3 \\
      \frac{1}{N-2} (\sum \limits_{j=1}^{N} x[n-j] \\
			  \ \ \ - \max \limits_{1\leq j \leq N} x[n-j] \\
			  \ \ \ - \min \limits_{1\leq j \leq N} x[n-j] ), & \text{otherwise}
    \end{cases}
\end{split}
\end{equation}
\ifdefined\AddSpace\vspace{3mm}\fi

\ifdefined\IEEEHeadingFormat
\vspace{2mm}
\textbf{4. Kalman Filtering Algorithm}
\vspace{2mm}

\else
\subsubsection{Kalman Filtering Algorithm} 
\fi
Kalman Filtering~\cite{kalman1960new} is one of the most well-known data fusion and estimation methods. One of its significant advantages is that it is the optimal estimator in systems with white Gaussian noise.

\ifdefined\TechReport
The algorithm we use is a one-dimensional Kalman Filter. Our terminology and notations are based on the standard literature, e.g.,~\cite{papoulis2002probability}.

\ifdefined\AddSpace\vspace{3mm}\fi
\begin{table}[!h]
		\centering
    \begin{tabular}{| l | p{6cm}|}
    \hline
	  $x[n]$ & The observed ETE of the $n^{th}$ sample. \\
	  $s[n]$ & The estimated ETE at $n$, given the measurements up to $n$. \\
	  $s[n|n-1]$ & The estimated ETE at $n$, given the measurements up to $n-1$. \\
	  $w[n]$ & The ETE signal noise of the $n^{th}$ sample. \\
	  $v[n]$ & The measurement noise of the $n^{th}$ sample. \\
	  $P[n]$ & The estimated variance of the ETE. \\
	  $P[n|n-1]$ & The estimated variance at $n$, given the measurements up to $n-1$. \\
	  $K[n]$ & The Kalman gain. \\
	  $W[n]$ & The estimated variance of $w[n]$, given the measurements up to $n-1$.  \\
	  $V[n]$ & The estimated variance of $v[n]$, given the measurements up to $n-1$.  \\
	  \hline
    \end{tabular}
    \caption{Kalman Filter Notations}
    \label{Notations}
\end{table}
\ifdefined\AddSpace\vspace{3mm}\fi

\textbf{Modeling the system.}
In the general Kalman Filtering model, the system equation is $s[n]=F \cdot s[n-1] + w[n]$, where F is the state transition coefficient. In our context the client does not have any information about how the ETE changes as a function of time, and therefore it is assumed that the state transition coefficient is 1. Hence, the Kalman system equation is given by~\ref{eq:system}.

\begin{equation}
\label{eq:system}
s[n]=s[n-1]+w[n]
\end{equation}
\ifdefined\AddSpace\vspace{3mm}\fi

The Kalman observation equation is given by:

\begin{equation}
\label{eq:observation}
x[n]=s[n]+v[n]
\end{equation}
\ifdefined\AddSpace\vspace{3mm}\fi

Based on the two equations above, we present the prediction equations and the update equations, which are the core of the Kalman Filtering algorithm.

\textbf{Prediction equations.} The client uses the prediction equations in step~1 of Fig.~\ref{fig:PredictApproach} to estimate the next ETE based on the first $n-1$ measurements.

\begin{equation}
s[n|n-1]=s[n-1]
\end{equation}

\begin{equation}
P[n|n-1]=P[n-1]+W[n]
\end{equation}
\ifdefined\AddSpace\vspace{3mm}\fi

\textbf{Update Equations.} The client uses the update equations in step~3 of Fig.~\ref{fig:PredictApproach} to update its state based on the new measurement, $x[n]$.

\begin{equation}
s[n]=s[n|n-1]+K[n](x[n]-s[n|n-1])
\end{equation}

\begin{equation}
K[n]=\frac{P[n|n-1]}{P[n|n-1]+V[n]}
\end{equation}

\begin{equation}
P[n]=(1-K[n])P[n|n-1]
\end{equation}
\ifdefined\AddSpace\vspace{3mm}\fi

\textbf{Variance estimation.}
$W[n]$ is defined to be the estimated variance of $w[n]$, and $V[n]$ is the estimated variance of $v[n]$. By Eq.~\ref{eq:system} and Eq.~\ref{eq:observation}, we have $w[n]=s[n]-s[n-1]$, and $v[n]=x[n]-s[n]$. Hence, the variance of $w[n]$ and $v[n]$ can be estimated by the \emph{sample variance} using the last $N$ values of $x[\cdot]$ and $s[\cdot]$, as follows:

\begin{equation}
\begin{split}
W[n] = \frac{1}{N} \cdot \sum \limits_{i=1}^{N} ( (s[n-i]-s[n-i-1])- \\ 
       (\frac{1}{N} \cdot \sum \limits_{j=1}^{N} (s[n-j]-s[n-j-1])))^2
\end{split}
\end{equation}

\begin{equation}
\begin{split}
V[n] = \frac{1}{N} \cdot \sum \limits_{i=1}^{N} ((x[n-i]-s[n-i])- \\
       (\frac{1}{N} \cdot \sum \limits_{j=1}^{N} (x[n-j]-s[n-j])))^2
\end{split}
\end{equation}

\else
\begin{sloppypar}
We model the ETE prediction problem as a one-dimensional Kalman Filtering problem, and derive the prediction equations and the update equations. We also propose variance estimation functions that fit our model. The details are presented in~\cite{OneClockTR}.
\end{sloppypar}
\fi

\begin{figure*}[htbp]
	\centering
  \begin{subfigure}[t]{.35\textwidth}
  \centering
  \fbox{\includegraphics[height=5.8\grafflecm]{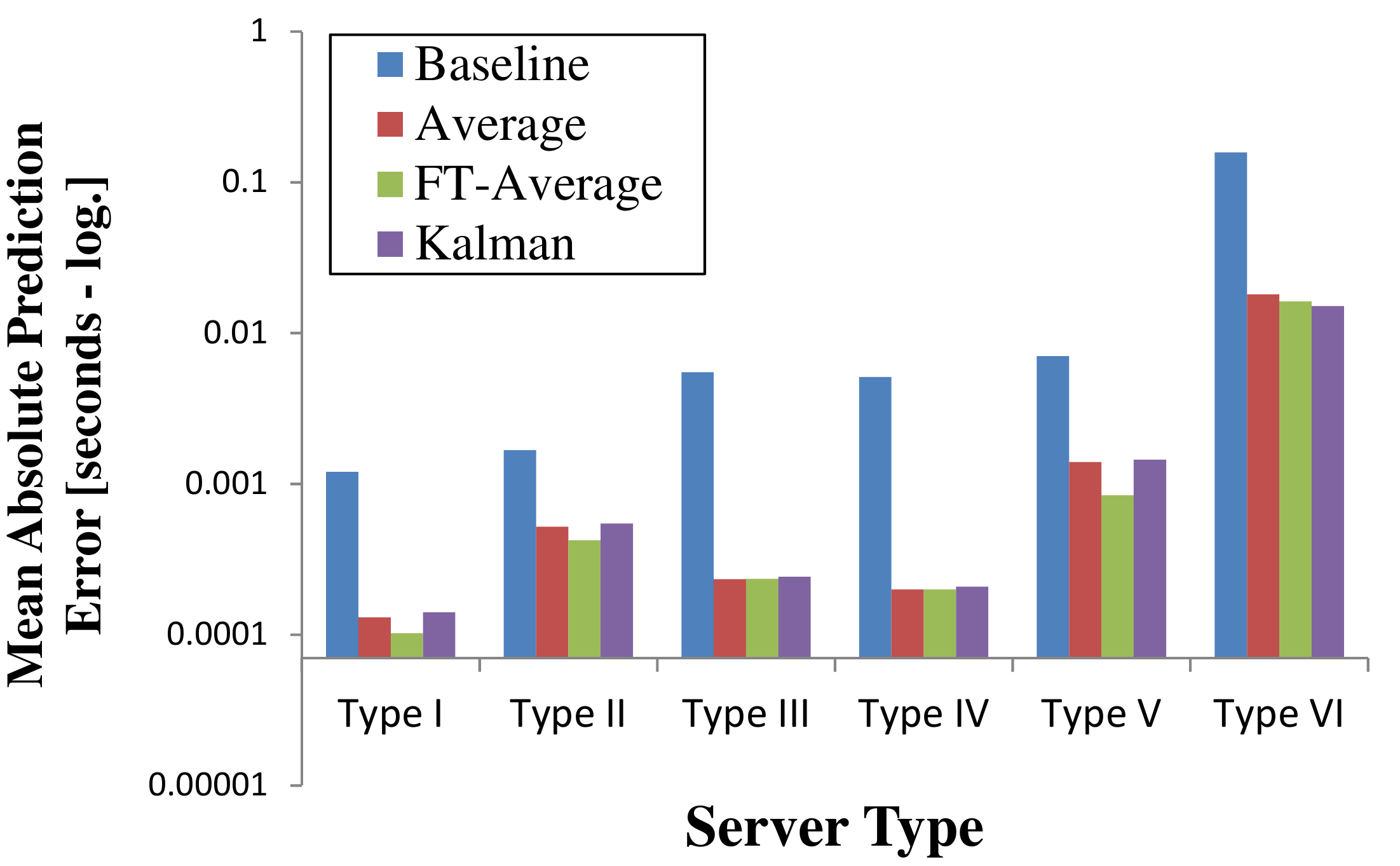}}
	\captionsetup{justification=centering}
  \caption{Performance on different platforms.}
  \label{fig:TypesGraph}
  \end{subfigure}%
  \begin{subfigure}[t]{.325\textwidth}
  \centering
  \fbox{\includegraphics[height=5.8\grafflecm]{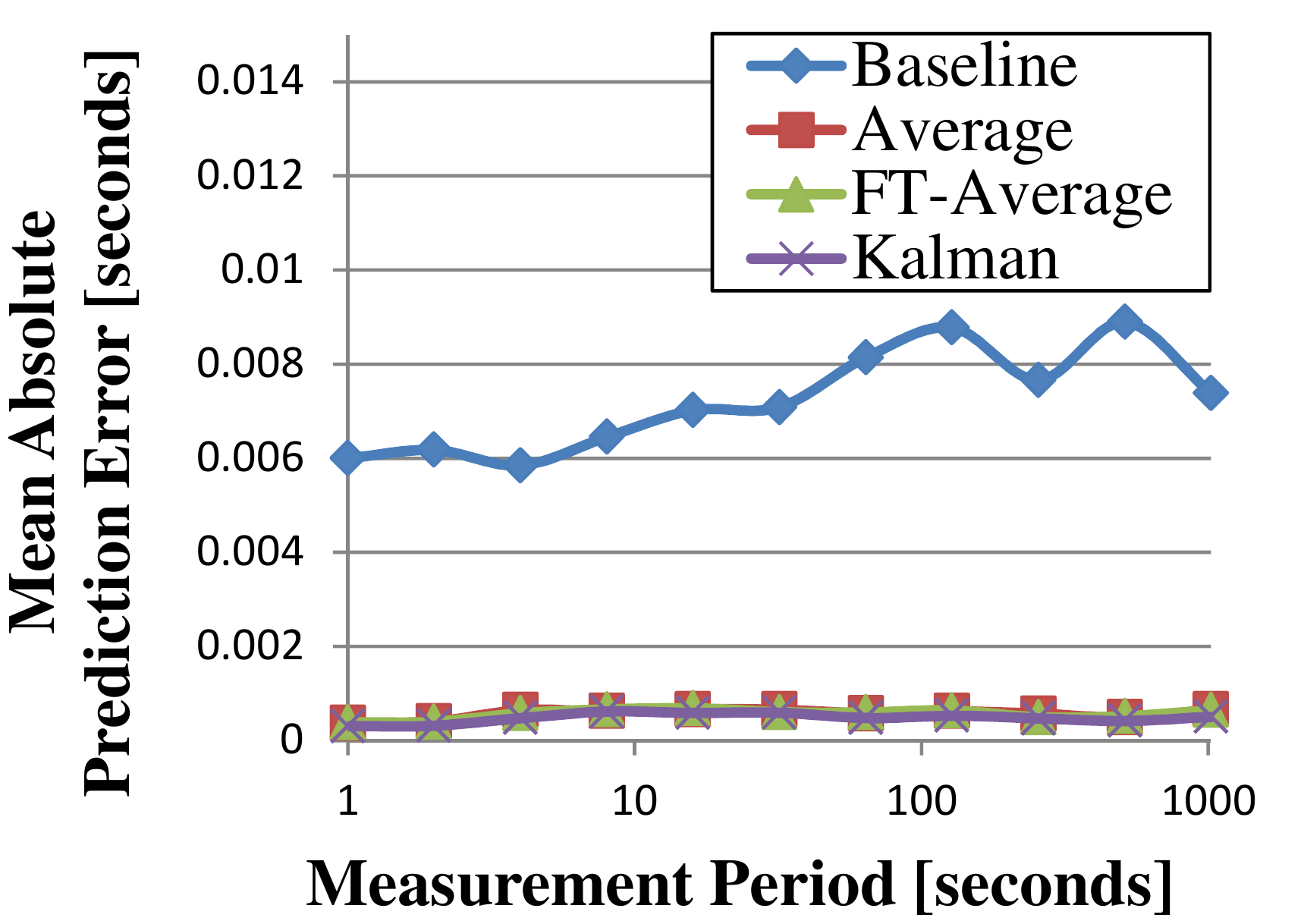}}
	\captionsetup{justification=centering}
  \caption{Periodic measurement.}
  \label{fig:PeriodGraph}
  \end{subfigure}%
  \begin{subfigure}[t]{.325\textwidth}
  \centering
  \fbox{\includegraphics[height=5.8\grafflecm]{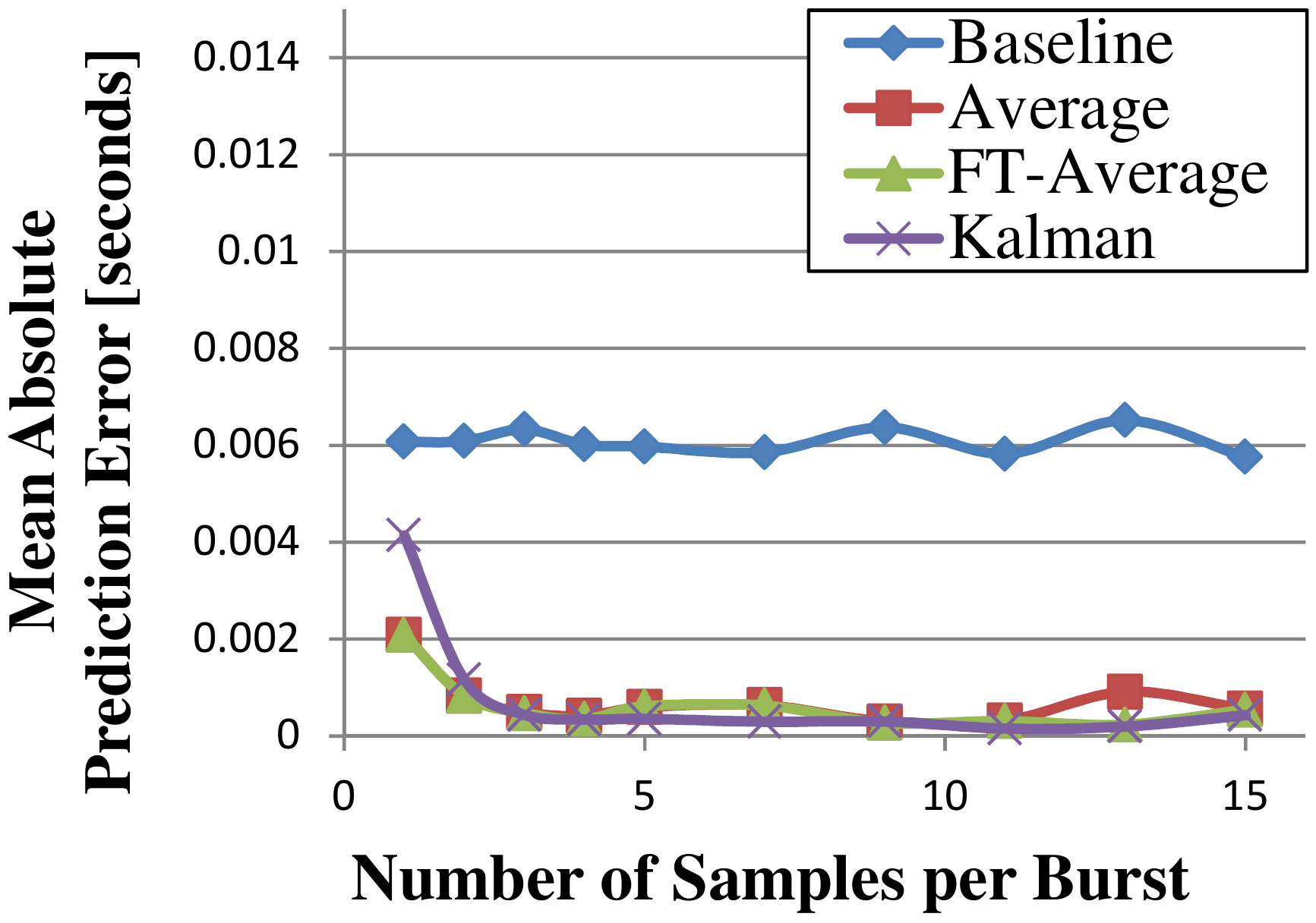}}
	\captionsetup{justification=centering}
  \caption{Bursty measurement.}
  \label{fig:BurstGraph}
  \end{subfigure}%
  \caption{Performance on various machine types (a). Type V machines were used in (b) and (c).}
  \label{fig:Evaluation}
	\ifdefined\CutSpace \vspace{-2mm} \fi
\end{figure*}

\begin{figure*}[htbp]
	\centering
  \begin{subfigure}[t]{.33\textwidth}
  \centering
  \fbox{\includegraphics[height=5.8\grafflecm]{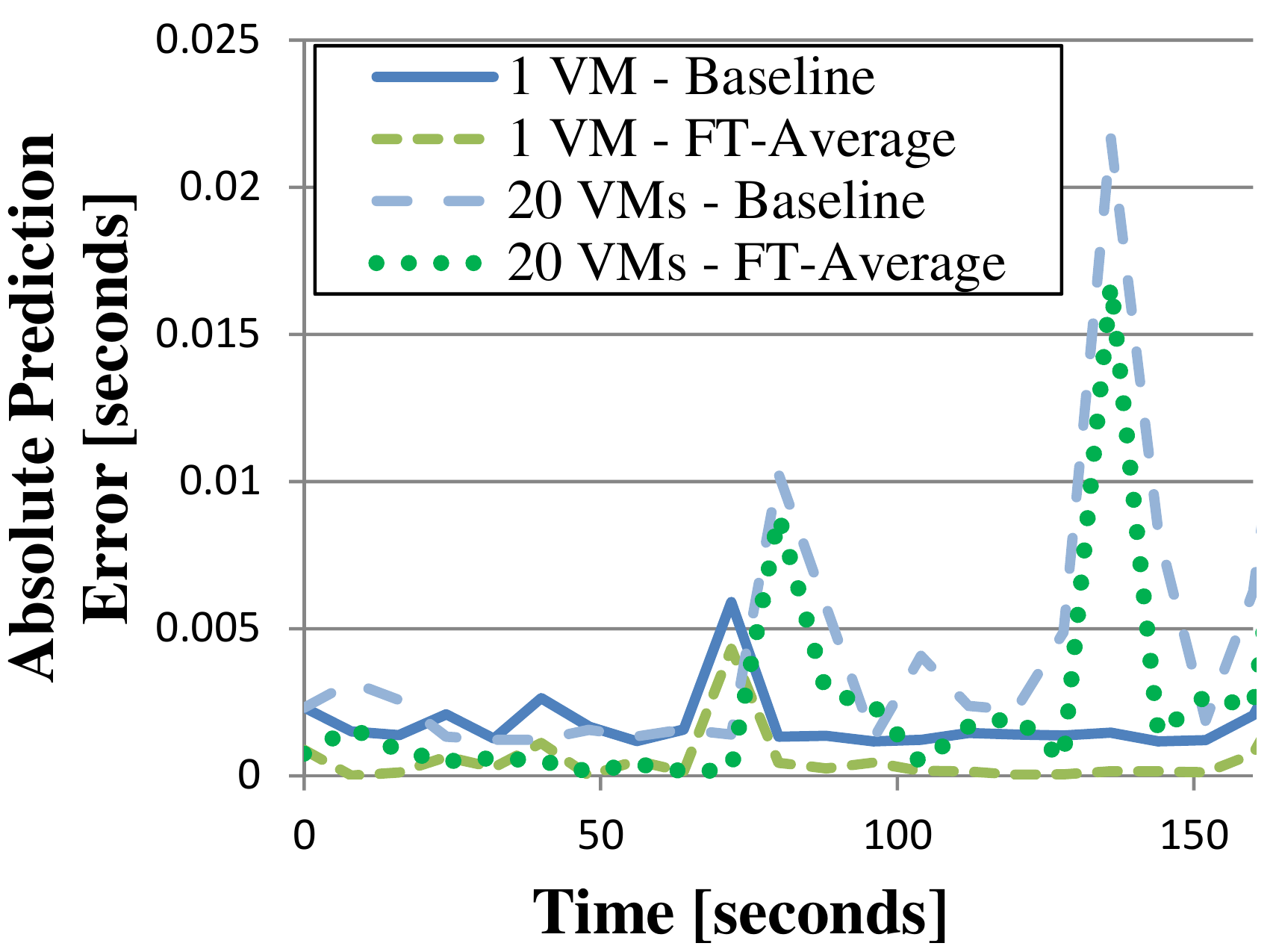}}
	\captionsetup{justification=centering}
  \caption{Performance on shared machine with 20 VMs, compared to machines with one VM.}
  \label{fig:VMGraph}
  \end{subfigure}%
  \begin{subfigure}[t]{.33\textwidth}
  \centering
  \fbox{\includegraphics[height=5.8\grafflecm]{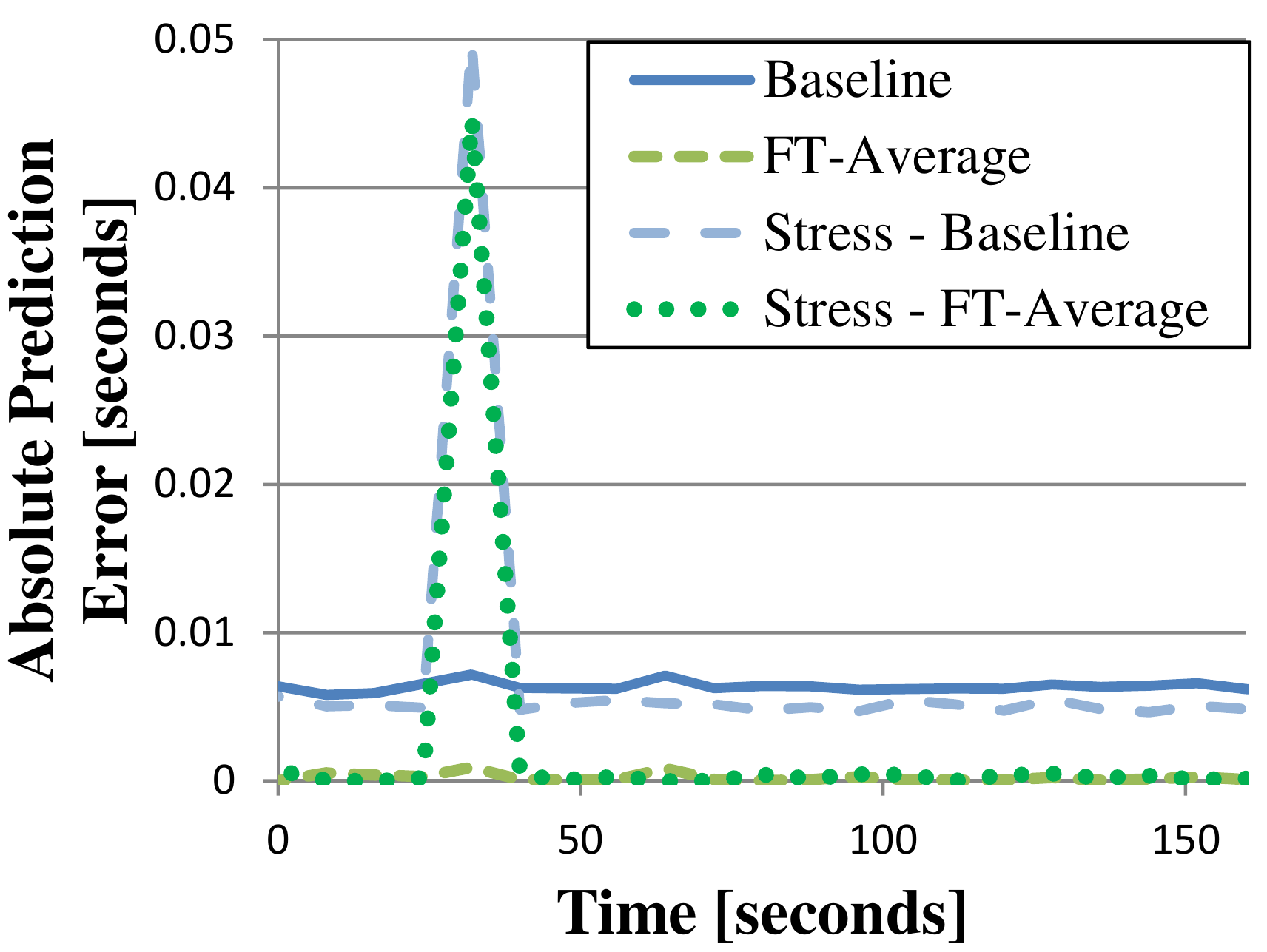}}
	\captionsetup{justification=centering}
  \caption{Performance on stressed server compared to unstressed server.}
  \label{fig:UtilGraph}
  \end{subfigure}%
  \begin{subfigure}[t]{.33\textwidth}
  \centering
  \fbox{\includegraphics[height=5.8\grafflecm]{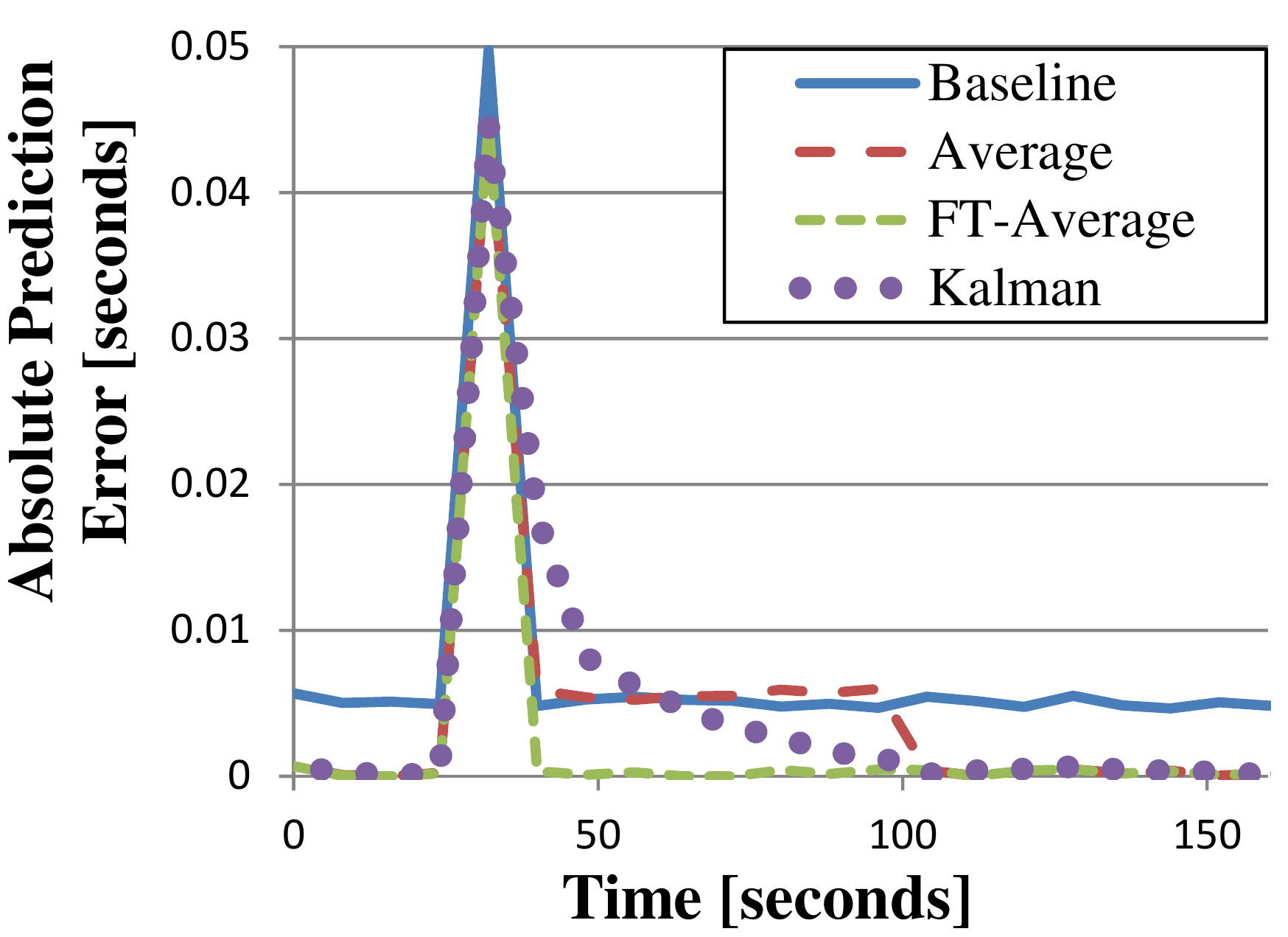}}
	\captionsetup{justification=centering}
  \caption{Occasional error spikes during a stress experiment.}
  \label{fig:SpikeGraph}
  \end{subfigure}%
  \caption{Instantaneous prediction error viewed over a $150$ second period. The behavior shows peaks under synthetic workload. (a)~was measured on Azure, and (b), (c) on Type V machines.}
  \label{fig:Stress}
	\ifdefined\CutSpace \vspace{-2mm} \fi
\end{figure*}

\ifdefined\AddSpace\vspace{5mm}\fi
\section{Evaluation}
\label{EvalSec}

\subsection{Background}
We implemented a prototype of the NETCONF time capability. The prototype was implemented as an extension to the OpenYuma~\cite{OpenYuma}, a NETCONF software implementation written in C over Linux. Our code is publicly available as open source~\cite{OneClockSource}.

\textbf{Goal.} The goal of the experiments was to evaluate our prediction-based scheduling approach over various machines, platforms, and under various workloads.

\textbf{Method.} We evaluated the three prediction algorithms (Sec.~\ref{PredictionSec}) on Linux-based servers in two academic testbeds, Emulab~\cite{EmulabProj} and DeterLab~\cite{DeterLabProj}, and in two public cloud platforms, Microsoft Azure~\cite{MicAzure}, and Amazon Web Services (AWS)~\cite{AmazonAWS}. Our measurements were performed on over 100 servers, for a total duration of over 5000 hours, summing up to over 3 million measurement samples.

\ifdefined\TechReport
Our results are based on measurements that were performed using a \emph{commit} RPC on the well-known \emph{toaster} YANG module~\cite{ToasterYANG}. In each experiment a NETCONF client sent scheduled RPC messages to a server, and the client recorded the $T_s$ and $T_e$ values. The experiments produced log files (at the client) containing $T_s$ and $T_e$ values, and then the three prediction algorithms were run offline.\footnote{In the current prototype we have not integrated the prediction algorithm logic into the NETCONF client. The prediction algorithms were run offline on the log files of the NETCONF client.} The three algorithms were run with $N=8$ in most of the runs\footnote{$N$ is the number of measurement samples used in each prediction computation. For further details see Sec.~\ref{PredictionSec}.}, except for specific runs in which the value of $N$ was different, as described below. 
\else
Further details about the evaluation method are presented in~\cite{OneClockTR}.
\fi

We quantify the accuracy of our prediction by observing the mean absolute prediction error. The prediction error of an RPC is defined as the difference between the predicted ETE and the measured ETE.

\ifdefined\AddSpace\vspace{3mm}\fi
\subsection{Experiment I: Performance on different platforms} 
In this experiment (Fig.~\ref{fig:TypesGraph}) we compared the prediction error of the three prediction algorithms on various server types. The prediction in this experiment was based on periodic sampling, with a measurement period of 8~seconds.\footnote{The measurement period is the elapsed time between two consecutive measurements.}
\ifdefined\TechReport
The list of servers we tested is presented in Table~\ref{MachineTypes}.
\else
The list of servers we tested is presented in~\cite{OneClockTR}. Types I~and~II were Virtual Machines (VM) running on AWS and Azure, respectively, and the other machine types were physical servers running on Emulab and DeterLab.
\fi

\ifdefined\TechReport
\ifdefined\AddSpace\vspace{3mm}\fi
\begin{table}[!h]
		\centering
    \begin{tabular}{|l|p{3.8cm}|p{2.5cm}|}
    \hline
    Type & Description & Platform / class \\ \hline \hline
    I & Public cloud (shared tenancy), 1GB memory & Amazon / t2.micro \\ \hline 
    II & Public cloud (shared tenancy), 768MB memory & Azure / A0 \\ \hline 
    III & Xeon E3 LP 2.4 GHz, 16GB memory & DeterLab / MicroCloud \\ \hline 
    IV & Xeon 2.1 GHz, 4GB memory & DeterLab / pc2133 \\ \hline 
    V & Quad Core Xeon E5530 2.4 GHz, 12GB memory & Emulab / d710 \\ \hline 
    VI & Dual Core Opteron 1.8 GHz, 4GB memory & DeterLab / bvx2200 \\ \hline 
    \end{tabular}
    \caption{Machine types.}
    \label{MachineTypes}
\end{table}
\ifdefined\AddSpace\vspace{3mm}\fi
\fi

As shown in Fig.~\ref{fig:TypesGraph}, the prediction algorithms significantly reduced the prediction error compared to the baseline approach. The experiment shows that in most of the cases FT-Average produces the lowest error.

\ifdefined\AddSpace\vspace{3mm}\fi
\subsection{Experiment II: Periodic vs. bursty measurement} 
We compared periodic measurement and burst-based measurement (see Fig.~\ref{fig:PeriodGraph} and~\ref{fig:BurstGraph}). The periodic measurement was performed at various measurement periods, and the burst measurement was performed with various burst sizes, and with a fixed period of one measurement per second. 
We note that in this experiment the error produced by the three algorithms is very similar.

Interestingly, the results show that a burst of 4 samples suffices to produce similar results to a periodic measurement.

In the periodic measurement (Fig.~\ref{fig:PeriodGraph}) the lowest prediction error was achieved with a period of one measurement per second. We were not able to test lower measurement periods due to a performance limitation in the NETCONF client we used.

\ifdefined\TechReport
Another interesting observation is that when the measurement period was on the order of one minute or more we observed slightly higher ETE values (Fig.~\ref{fig:PeriodGraph}) than when the measurement period was on the order of a few seconds. This can be explained by the server's cache policy, which allows better performance for operations that are performed frequently. 

\fi

\ifdefined\AddSpace\vspace{3mm}\fi
\subsection{Experiment III: Performance under synthetic workload}  
In this experiment we studied the prediction error in stressed NETCONF servers, compared to the error in unstressed NETCONF servers. We used two methods to stress the machines: (i) We used the \emph{lookbusy}~\cite{Lookbusy} utility to inject synthetic workload (Fig.~\ref{fig:UtilGraph}). We configured the utility to run at a CPU utilization of 95\% and at a memory utilization of 95\%. (ii) We used the Azure platform to run multiple VMs on the same physical machine (Fig.~\ref{fig:VMGraph}). We ran 20 VMs on the same machine, where one of the VMs was the NETCONF server.

During the stress experiments we observed that most of the ETE measurements were unaffected by the stress, but as depicted in Fig.~\ref{fig:VMGraph} and~\ref{fig:UtilGraph}, there were occasional spikes in the ETE, causing temporary high prediction error. As shown in Fig.~\ref{fig:VMGraph} and~\ref{fig:UtilGraph}, prediction error of the FT-Average algorithm during the ETE spikes is slightly lower than the baseline error, and during most of the run the prediction error of the FT-Average is \textbf{significantly} lower than the baseline error.

Fig.~\ref{fig:SpikeGraph} compares the three prediction approaches during an ETE spike. As illustrated in Fig.~\ref{fig:SpikeGraph}, the FT-Average algorithm was the most resilient to these spikes, as it ignores the maximal and minimal measurement samples, and thus ignores the peak ETE value. As depicted in the figure, the two other algorithms were more sensitive to these spikes.

\ifdefined\AddSpace\vspace{5mm}\fi
\section{Discussion}

\textbf{Prediction method.} 
As discussed in the previous sections, we analyzed three prediction algorithms.
Kalman filtering was used as it is one of the most celebrated and popular data fusion algorithms. The Average approach was chosen due to its simplicity, and FT-Average due to its resilience to occasional isolated noisy measurements. 

The experimental results show that the prediction error offered by the three algorithms is similar, and is significantly lower than the baseline error. The FT-Average approach showed slightly lower prediction error in most of the experiments. FT-Average is especially advantageous in the presence of occasional spikes in the ETE, as it inherently ignores the erroneous measurement. Interestingly, even a short burst of $N=4$ measurements allows the simple FT-Average algorithm to predict the ETE with a very low prediction error.

\textbf{Measurement period.} 
The measurement period is the elapsed time between two consecutive measurements. Frequent measurements may be more sensitive to changes in the ETE, and allow a more accurate prediction. On the other hand, if measurements are performed too frequently they may affect the server's performance. Since the client continuously monitors the prediction error, the client can dynamically change the measurement period for each server to improve the prediction error. Thus, an interesting extension to our work would be to implement an algorithm that dynamically changes the measurement period for each server.



\textbf{RESTCONF.} 
An interesting next step would be to extend the scope of our work, and apply it to the emerging RESTCONF~\cite{restconf}.
This work would be especially interesting in the context of resource constrained servers, such as IoT devices.

\ifdefined\TechReport
\textbf{Multiple RPC types.}
The ETE of an RPC depends on the RPC type. Thus, the prediction method should be used on a per-RPC-type basis. A possible extension of our work would be to consider how to predict the ETE of RPC type A using measurements of RPC type B.

\textbf{Time zone issues.}
Since the client and servers may be spread across multiple time zones. The NETCONF date-and-time format specifies the time zone for each timestamp, thereby avoiding ambiguity in the timestamp value. Moreover, to avoid problems that may arise during Daylight Saving Time (DST) changes, the client can invoke scheduled RPCs using the UTC time zone, which is not subject to DST changes.

\textbf{Impact of the network delay on the measurements.}
When a client sends a scheduled RPC message, the message must be sent in advance, allowing the message to arrive to the server before the scheduled time. Thus, as the network delay increases, the client must send the scheduled RPC sooner. In our experiments we considered the network delay when planning the time at which RPCs are sent. Note that the ETE is a metric of the servers' performance, and is not affected by the network delay.

\textbf{Inter-RPC influence.} The ETE of an RPC may be affected by other RPCs that are running in parallel, or are scheduled to run in parallel. The prediction approach presented in this paper estimates the RPC's execution time, given that the server is subject to workload by other tasks or other RPCs running in parallel. In our evaluation we mimicked these scenarios by synthetically creating additional workload on the servers.

\fi

\ifdefined\AddSpace\vspace{2.5mm}\fi
\section{Conclusion}
\onec\ is a generic approach for using accurate time in distributed applications.
As NETCONF is gaining momentum and penetrating various new network applications, \onec\ seems like a natural extension that can add the time dimension to network configuration and management.

We analyzed three prediction algorithms, and found the simple FT-Average to be the most accurate algorithm in most of the experiments.
Our experimental evaluation confirms that prediction-based scheduling provides a high degree of accuracy in various diverse environments, decreasing the prediction error by an order of magnitude compared to the na\"{\i}ve baseline approach. 

\section{Acknowledgments}
We gratefully acknowledge Alon Schneider and Eylon Egozi for their help with the prototype implementation. We thank Amazon for supporting our AWS experiments by an AWS in Education Research Grant award. We gratefully acknowledge the Emulab project~\cite{EmulabProj} and the DeterLab project~\cite{DeterLabProj} for the opportunity to perform our experiments on their testbeds. This work was supported in part by the ISF grant 1520/11.

\ifdefined\AddSpace\vspace{5mm}\fi

\bibliographystyle{IEEEtran}
\bibliography{time}
\label{last-page}

\end{document}